\documentclass[aps,twocolumn,preprintnumbers,groupedaddress]{revtex4-1}
\pdfoutput=1

\usepackage{tabu}
\usepackage[usenames,dvipsnames,table]{xcolor}
\usepackage{graphicx,amsmath,amssymb,amsthm,multirow,array,bm,bbm,esint}
\usepackage[mathscr]{eucal}
\usepackage[bbgreekl]{mathbbol}
\usepackage{epsf,amsfonts}
\usepackage{physics}
\usepackage{dsfont, mathtools}
\usepackage{slashed}
\usepackage{hyperref}
\usepackage{mathdots}

\usepackage{ulem}

\hypersetup{
    pdfstartview={FitH},    
    pdftitle={Diagnosing collisions in the interior of a wormhole},    
    pdfauthor={Felix Haehl and Ying Zhao},     
    colorlinks=true,       
    linkcolor=blue,          
    citecolor=red,        
    filecolor=magenta,      
    urlcolor=blue           
}

\usepackage{upgreek}

\usepackage{tikz}

\begin{document}

\title{Diagnosing collisions in the interior of a wormhole}

\author{Felix M.\ Haehl}
\author{Ying Zhao}

\affiliation{School of Natural Sciences, Institute for Advanced Study\\
1 Einstein Drive, Princeton, NJ 08540, USA.}

\begin{abstract}
Two distant black holes can be connected in the interior through a wormhole. Such a wormhole has been interpreted as an entangled state shared between two exterior regions. If Alice and Bob send signals into each of the black holes, they can meet in the interior. In this letter, we interpret this meeting in terms of the quantum circuit that prepares the entangled state: Alice and Bob sending signals creates growing perturbations in the circuit, whose overlap represents their meeting inside the wormhole. We argue that such overlap in the circuit is quantified by a particular six-point correlation function. Therefore, exterior observers in possession of the entangled qubits can use this correlation function to diagnose the collision in the interior without having to jump in themselves.
\end{abstract}


 \maketitle

\section{Introduction}
Two distant black holes can be connected in the interior through a wormhole. In the context of gauge/gravity duality \cite{Maldacena:1997re}, such an Einstein-Rosen (ER) bridge can be interpreted as an entangled state between two asymptotic quantum field theories \cite{Maldacena:2001kr}: Alice and Bob in the two asymptotic regions share entangled EPR pairs. At the heart of this ``ER = EPR" mechanism \cite{Maldacena:2013xja} lies the following mystery: even though Alice's and Bob's systems are not coupled, they can send signals into the dual wormhole at appropriate times in a way such that they meet in the interior \cite{Marolf:2012xe}.

In this letter, we explain the ``meeting'' of two signals in the interior in terms of a quantum circuit. The circuit consists of gates acting on the shared EPR pairs, thus encoding the dynamical evolution of the geometry. More precisely, the quantum circuit prepares the entangled state corresponding to the wormhole at any given time \cite{Hartman:2013qma,Susskind:2014moa}. When Alice and Bob send signals into the bulk geometry, they create perturbations in the circuit, which initially grow exponentially. Depending on the time when the signals are sent in, the two perturbations can have overlap in the quantum circuit. We argue that this overlap is the ``EPR manifestation'' of the meeting of the two signals in the interior.

We offer a precise quantitative measure of the overlap by defining a notion of size for the operator used to create the perturbations. This size can be expressed as a particular six-point correlation function. On the gravity side, this correlation function detects the backreaction on the geometry caused by the high-energy collision in the interior, and hence can diagnose the meeting of two signals from different boundaries. This demonstrates that ``ER = EPR'' can be treated as a quantitative and dynamical statement: it can be used to diagnose processes in the wormhole interior from the safe outside without jumping in.

\section{A meeting in the interior of the wormhole}

\begin{figure}
\begin{center}
 \includegraphics[width=\columnwidth]{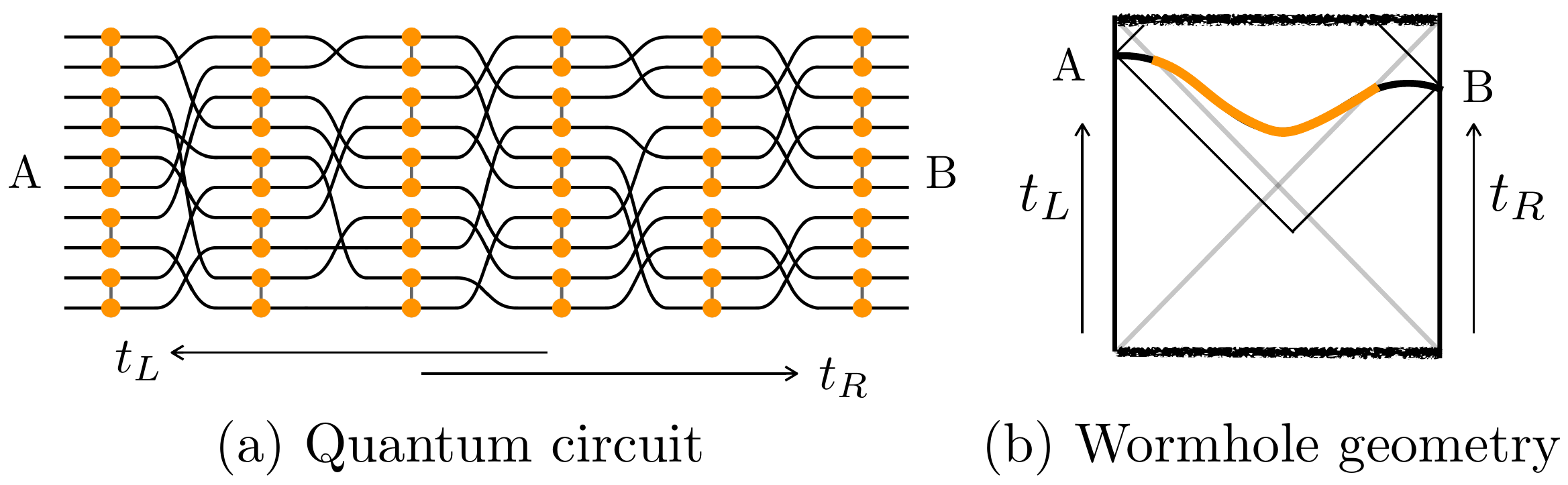}
\end{center}\vspace{-.5cm}
\caption{We model the dynamical evolution of the entangled thermofield double state by a simple quantum circuit: at each time step the shared qubits (black lines) are randomly grouped into $\frac{S}{2}$ pairs, and on each pair a randomly chosen 2-qubit gate (orange dots) is applied.}
\label{thermofield_double_time_evolved}
\end{figure}

The eternal black hole geometry (Figure \ref{thermofield_double_time_evolved}b) implies that objects falling in from the two disconnected boundaries can meet in the shared interior. In this section we will give a quantum circuit interpretation of this phenomenon in the spirit of the ``ER = EPR'' paradigm \cite{Zhao:2020gxq}.
Let us begin with the unperturbed thermofield double state shared between Alice and Bob:
\begin{equation}
\label{eq:TFDdef}
  |\text{TFD} \rangle = \sum_k e^{-\frac{\beta}{2} E_k} \, |E_k \rangle_L \otimes |E_k\rangle_R \,,
\end{equation}
where $|E_k\rangle$ are the energy eigenstates of the individual boundary systems at inverse temperature $\beta$.
We model this state by $S$ maximally entangled EPR pairs. As the left (right) boundary time increases, the circuit grows toward the left (right), as does the interior of the wormhole \cite{Susskind:2014moa,Stanford:2014jda}. We identify the circuit time with the right boundary time $t_R$ and minus of the left boundary time $-t_L$ \footnote{The minus sign appears here because the left boundary time is in the opposite direction with the Schwarschild time.} (see Figure \ref{thermofield_double_time_evolved}).

\begin{figure}
\begin{center} 
 \includegraphics[width=\columnwidth]{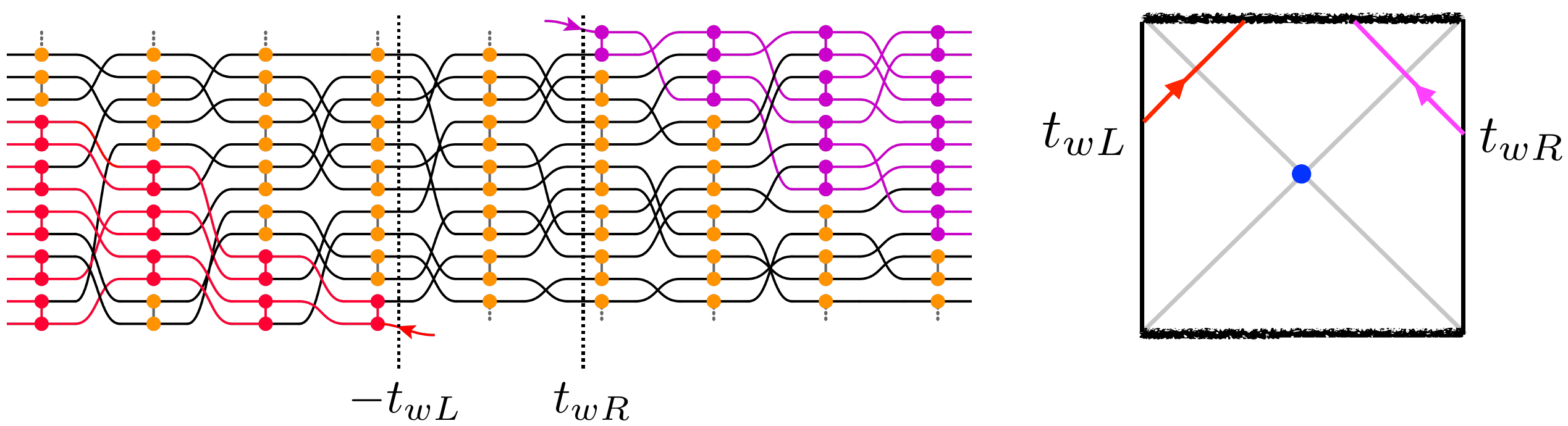}
\end{center}\vspace{-.5cm}
\caption{{\it Left:} The red (pink) arrow represents the extra qubit due to Alice's (Bob's) perturbation. Any qubits that interact directly or indirectly with these perturbations, get perturbed relative to the original circuit describing the thermofield double state. When $t_{wL}+t_{wR}>0$, the two perturbations do not have overlap in the quantum circuit. {\it Right:} Correspondingly, the signals sent into the bulk hit the singularity before they have a chance to meet inside the wormhole.}
\label{circuit_40}
\end{figure}

Alice sends in one thermal-scale quantum from the left boundary at time $t_{wL}$. In the quantum circuit, an extra qubit enters at circuit time $-t_{wL}$ and propagates toward the left. This corresponds to the red line in Figure \ref{circuit_40}. Bob creates a similar perturbation at the right boundary at time $t_{wR}$. This creates another perturbation in the quantum circuit that grows toward the right (shown in pink). Depending on the relative ordering of $-t_{wL}$ and $t_{wR}$, the two perturbations may or may not have an interval of overlap.

\begin{figure}
 \begin{center}
  \includegraphics[width=\columnwidth]{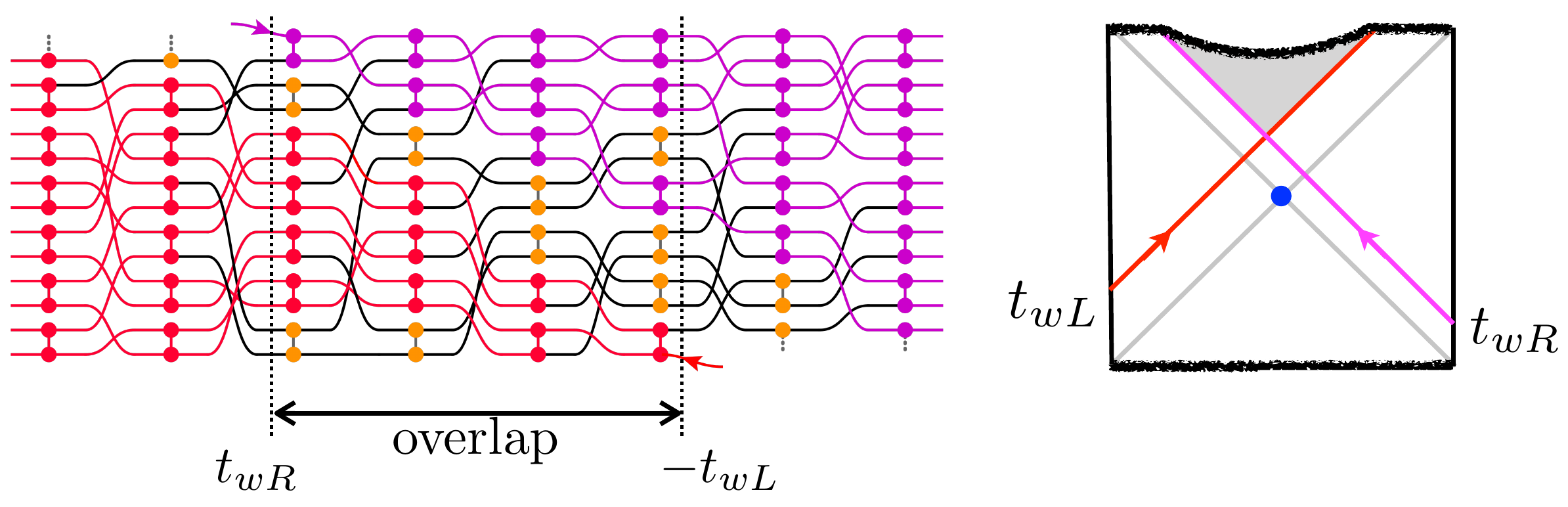}
 \end{center}\vspace{-.5cm}
 \caption{{\it Left:} When $t_{wL}+t_{wR}<0$, the two perturbations have overlap in the quantum circuit. {\it Right:} In the bulk geometry this corresponds to a collision inside the wormhole.}
\label{circuit_5}
\end{figure}

When $t_{wL}+t_{wR}>0$, there is no overlap between the two perturbations in the circuit. Correspondingly, the two perturbations do not meet in the interior (Figure \ref{circuit_40}). On the other hand, when $t_{wL}+t_{wR}<0$, the overlap of the two perturbations in the quantum circuit represents the meeting of the two signals in the interior geometry (Figure \ref{circuit_5}). The larger the overlap is, the stronger the collision is. 

\section{Diagnosing the collision in the wormhole interior}

Having explained the qualitative picture, we will now describe how to detect the interior collision. 

We want to quantify the overlap in the quantum circuit in Figure \ref{circuit_5}. Imagine experimentalists Alice and Bob try to build the quantum state in their laboratory. They start from some shared EPR pairs and implement the following circuits:
\begin{equation*}
      \includegraphics[width=\columnwidth]{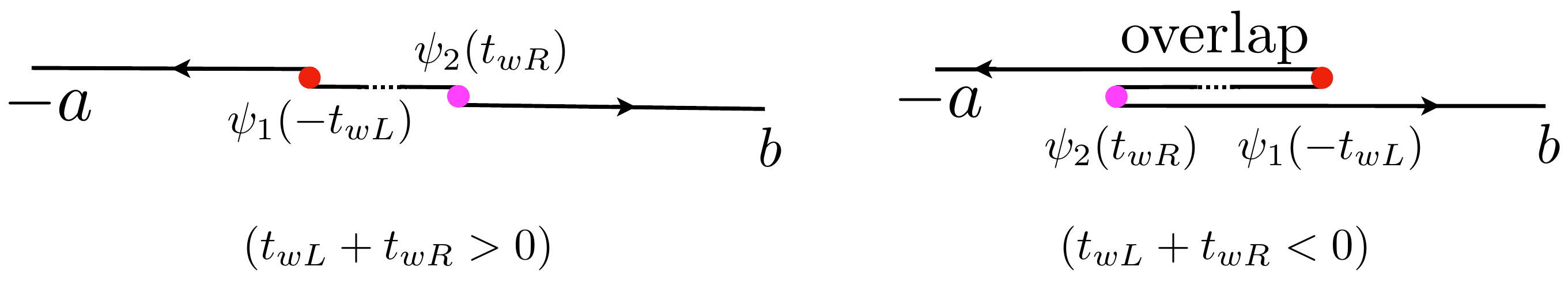}
\end{equation*}
The dotted region indicates the starting point for both. In the first case, when the perturbations never overlap, Alice can proceed toward the left and insert the red perturbation $\psi_1$, while Bob will proceed toward the right and insert the pink perturbation $\psi_2$. In the second case, when the perturbations have overlap, before inserting their extra qubits both Alice and Bob have to do {\it backward} time evolution.\footnote{Note that we do not mean Alice and Bob themselves travel backward in time. 
They merely need to reverse the Hamiltonian acting on the qubits in their lab.
This is not only possible but also practical in quantum laboratories \cite{Swingle:2016var}.} This makes the procedure complex.
At the end there is a timefold in the quantum circuit. As the fold grows, the collision gets stronger. In order to diagnose it, we think of the entire circuit as a single (complicated) operator preparing a state, and we quantify the ``size'' of this operator.

Size of an operator is defined as the average number of fundamental constituents it is made of. For example, in Feynman's parton theory of hadrons, the number of ``wee partons" increases as a particle's momentum increases \cite{Kogut:1972di}. In string theory the length of a string increases as we observe it with higher and higher resolution \cite{Karliner:1988hd,Susskind:1993aa}. In the Sachdev-Ye-Kitaev (SYK) model the size of an operator is defined as the average number of fermions making up the operator \cite{Roberts:2018mnp,Qi:2018bje}. The growth of operator size was also argued to explain the gravitational force \cite{Susskind:2017ney,Susskind:2018tei}. Let us explore this connection further and relate the collision in the wormhole interior to the size growth of a suitable operator.

The size growth of one single perturbation operator in the quantum circuit was well studied \cite{Susskind:2014jwa,Roberts:2018mnp,Qi:2018bje}.  For concreteness, consider the SYK model with $N$ Majorana fermions $\psi_j$. The infinite temperature size $n_\infty[{\cal O}]$ of some operator ${\cal O}$ can be defined as the average number of fermions making up the operator \cite{Roberts:2018mnp}:
\begin{equation}
  n_\infty[{\cal O}] \equiv \frac{1}{4} \sum_j \text{tr} \left( \{ {\cal O} ,\psi_j \}^\dagger\, \{{\cal O},\psi_j \} \right) \,.
\end{equation}
This notion of infinite temperature operator size can be generalized to a measure of size in thermal states \cite{Qi:2018bje}:
\begin{equation}
\label{eq:nBetaDef}
 \frac{n_\beta[{\cal O}]}{n_{max}} \equiv \frac{n_\infty[{\cal O}\rho^{\frac{1}{2}}] - n_\infty[\rho^{\frac{1}{2}}] }{n_{max} - n_\infty[\rho^{\frac{1}{2}}]} \,,
\end{equation}
where $n_{max} = \frac{N}{2}$ is the size of a completely scrambled operator. The thermal size \eqref{eq:nBetaDef} of a single fermion $n_\beta[\psi_1(t)]$ can be related to an out-of-time-ordered four-point function \cite{Qi:2018bje}, which grows exponentially until it saturates to $n_{max}$ at a time of order $t \gtrsim t_*$:
\begin{equation}
    \label{eq:4point}
    \begin{split}
    \mathcal{F}_4(t) &\equiv 1-\frac{n_{\beta}[\psi_1(t)]}{n_{max}} \\
    &=  -\frac{\sum_j\tr(\psi_1(t)\psi_j\psi_1(t)\rho^{\frac{1}{2}}\psi_j\rho^{\frac{1}{2}})}{\sum_j\tr(\psi_1(t)\psi_1(t)\rho)\tr(\rho^{\frac{1}{2}}\psi_j\rho^{\frac{1}{2}}\psi_j)}\,.
\end{split}
\end{equation}

In our case, the operator that builds the circuit described above is composed of two perturbations:
\begin{equation}
e^{iHa}\psi_1(-t_{wL})\rho^{\frac{1}{2}}\psi_2(t_{wR})e^{iHb}
\end{equation} 
where $\rho^{\frac{1}{2}}$ represents the thermofield double state.
This operator summarizes the procedure of perturbing the thermofield double from the right side by $\psi_2$ at time $t_{wR}$ and from the left by $\psi_1$ at time $t_{wL}$. Taking $a$, $b$ large and positive implies that we are looking at the state at a late time. In order to detect the presence of the fold in the circuit, we define the renormalized size of the operator with two perturbations as follows \cite{Qi:2018bje,Haehl:2021sib}:
\begin{widetext}
\begin{align}
\label{size_detect}
	&\frac{n_{\text{ren}}}{n_{\text{max}}}\equiv\frac{n_{\infty}[e^{iHa}\psi_1(-t_{wL})\rho^{\frac{1}{2}}\psi_2(t_{wR})e^{iHb}]-n_{\infty}[e^{iHa}\rho^{\frac{1}{2}}e^{iHb}]}{n_{\text{max}}-n_{\infty}[e^{iHa}\rho^{\frac{1}{2}}e^{iHb}]}.
\end{align}
\end{widetext}

We can give a pictorial representation of the quantity \eqref{size_detect} for the two cases:
\begin{equation*}
\includegraphics[width=\columnwidth]{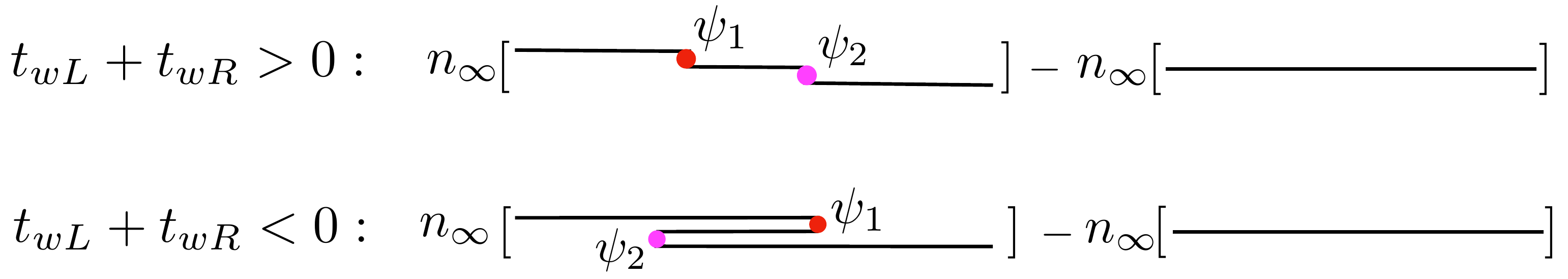}
\end{equation*}
Without the fold the renormalized size does not grow.
In the case where $t_{wR}+t_{wL}<0$, the fold makes a non-trivial contribution to the renormalized size. This contribution grows as the fold grows.

Our proposal is to use the renormalized size \eqref{size_detect} to diagnose the collision in the wormhole interior.

\begin{figure}
 \begin{center}                  
      \includegraphics[width=1.6in]{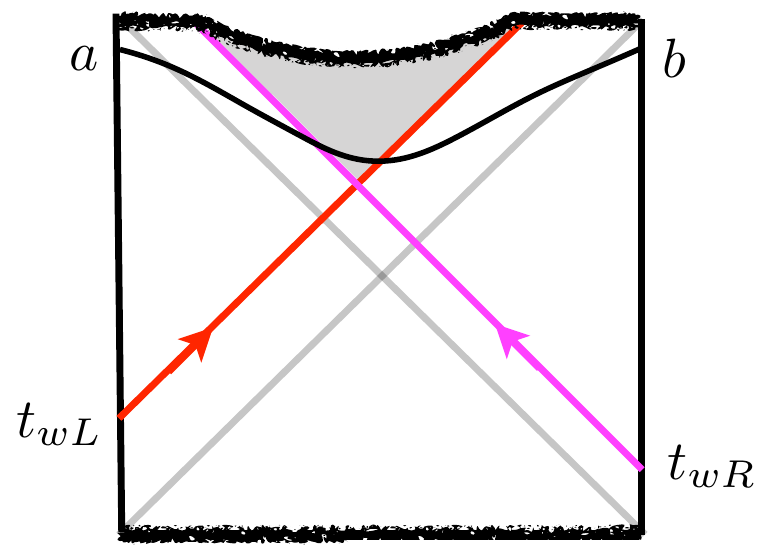}
      \caption{We consider the perturbed state at left and right times $a$ and $b$. The correlator \eqref{correlator} can be approximated by the bulk geodesic distance between these anchor points.}
  \label{Penrose_3}
  \end{center}
\end{figure}

Just as the size of a single perturbation operator is related to an out-of-time-ordered four-point function (c.f.\eqref{eq:4point}), the renormalized size we used in \eqref{size_detect} can be written as a six-point correlation function: $\frac{n_{\text{ren}}}{n_{\text{max}}} = 1-\mathcal{F}_6$, where
\begin{widetext}
 \begin{align}
 \label{correlator}
	\mathcal{F}_6	=\ &\frac{\sum_{j = 1}^N\Big(\bra{\text{TFD}}\psi_1^L(t_{wL})\psi_2^R(t_{wR})\Big)\ \psi_j^L(a)\psi_j^R(b)\ \Big(\psi_1^L(t_{wL})\psi_2^R(t_{wR})\ket{\text{TFD}}\Big)}{\sum_{j = 1}^N\bra{\text{TFD}}\psi_j^L(a)\psi_j^R(b)\ket{\text{TFD}}}\, . 
 \end{align}
 \end{widetext}
  This quantity has a simple interpretation in gravity. Consider the thermofield double perturbed by $\psi_1$  and $\psi_2$: $\psi_1^L(t_{wL})\psi_2^R(t_{wR})\ket{\text{TFD}}$. Now consider this state on a time slice anchored at large left and right times $a$ and $b$ (Figure \ref{Penrose_3}). If a collision has occurred, it will disrupt correlations between the left and right systems at these times. Therefore, the simplest measures of this process are left-right correlation functions between identical basis operators $\psi_j$. We sum over {\it all} fundamental fields $\psi_j$ in the theory to account for the universality of gravity: disruptions of correlations can occur between any of the degrees of freedom.

\begin{figure}
 \begin{center}                      
      \includegraphics[width=\columnwidth]{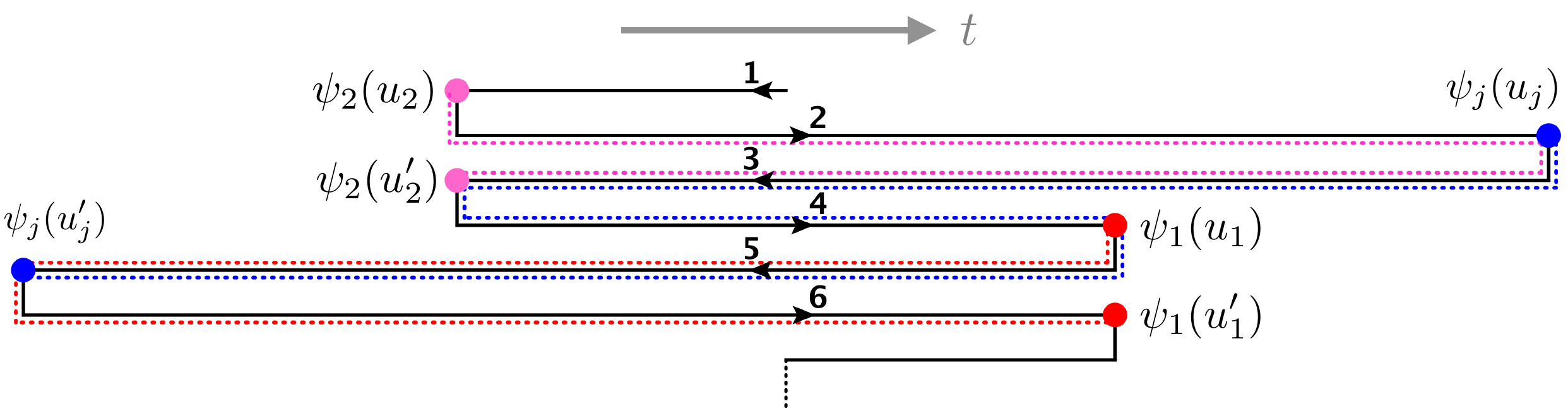}
      \caption{The complex time contour representing the correlator \eqref{correlator}. Real time goes towards the right.}
  \label{fold}
  \end{center}
\end{figure}

A contour representation of the correlator \eqref{correlator} is given in figure \ref{fold}, where the vertical (horizontal) direction represents Euclidean (Lorentzian) time. Notice that we need at least six `timefolds' to represent this correlator, which means that the configuration is `maximally out-of-time-order' \cite{Haehl:2017qfl,Haehl:2017pak}.

\section{Calculation}
In this section we will outline the calculation of gravitational contributions to the six-point function \eqref{correlator}. Our method follows a similar computation of the four-point function in \cite{Maldacena:2016upp} (see also \cite{Shenker:2014cwa,Maldacena:2017axo}).

To perform this calculation, for simplicity we consider a $(0+1)$-dimensional theory whose dynamical variable consists of time reparametrizations $u\mapsto t(u)$. This also arises as the asymptotic symmetry of dilaton gravity in $AdS_2$. Such `conformal' symmetry is generally broken by the choice of state. The {\it Schwarzian theory} additionally describes the leading explicit symmetry breaking effects of any such `nearly conformal field theory' (such as the low energy sector of SYK model \cite{Kitaev:2015talk,Maldacena:2016hyu}), as well as the boundary dynamics of $AdS_2$ gravity \cite{Maldacena:2016upp}. We will thus compute the correlator \eqref{correlator} by performing a path integral over time reparametrizations Goldstone modes $t(u)$ weighted by the Schwarzian action $S[t(u)] = -C \int du \, \{ t(u),u\}$:
\begin{equation}
\label{eq:F6SchwGen}
  {\cal F}_6 = {\cal N} \int [{\cal D}t] \, e^{i S[{t}({u})]} \, G_{\Delta_1}({u}_1,{u}_1') \, G_{\Delta_2}({u}_2,{u}_2') \, G_{\Delta_j} ({u}_j,{u}_j') \,.
\end{equation}
The bilocal operators $G_{\Delta}$ describe the coupling of the operator pairs to the reparametrization mode, which is obtained via a conformal transformation of the two-point function:
\begin{align}
    G_{\Delta}(u, u') =\qty(-\frac{t'(u)t'(u')}{(t(u)-t(u'))^2})^{\Delta} 
\end{align}
The normalization ${\cal N}$ is given by products of two-point functions.
The insertion points for our setup (figure \ref{fold}) are
\begin{align*}
   & u_1 = -t_{wL}-i\pi +i\delta_1,\ \  u_1' = -t_{wL}-i\pi-i\delta_1\\
    & u_2 = t_{wR}+i\delta_2,\ \qquad\quad\; u_2' = t_{wR}-i\delta_2\\
    &u_j = b,\ \qquad\qquad\qquad\quad u_j' = -a-i\pi
\end{align*}
where we introduced small Euclidean smearing $\delta_{1,2}$ to regulate. 

In order to perform the integral over reparametrizations, we make the following observations: because the bulk geometry is always exactly $AdS_2$, the effect of local operator insertions is to `kick' the boundary trajectory in a different direction. More formally, the operators source shock waves along which $AdS_2$ black hole solutions get glued together \cite{Maldacena:2016upp,Goel:2018ubv}. Both sides of the shock wave have a definite $SL(2,\mathbb{R})$ charge, and their difference compensates for the charge due to the matter fields. All the information about the geometry is therefore captured by $SL(2,\mathbb{R})$ gauge transformations at the operator insertion points. This reduces the path integral \eqref{eq:F6SchwGen} to an integral over suitable $SL(2,\mathbb{R})$ gauge parameters.

Restricting to exponentially growing modes, the piecewise $SL(2,\mathbb{R})$ transformations along the contour in figure \ref{fold} takes the following form:
\begin{widetext}
\begin{equation}
\label{eq:t6Def}
\begin{split}
  t(u) &= x - \frac{(1+x)^2X^-}{2+(1+x)X^-} \, \theta(2,3) + \frac{(1-x)^2X^+}{2+(1-x)X^+} \, \theta(5,6)+ \frac{(1-x)^2Y^+}{2+(1-x)Y^+} \, \theta(3)+ (\cdots) \, \theta(4) - \frac{(1+x)^2Y^-}{2+(1+x)Y^-} \, \theta(5) \,,
 \end{split}
 \end{equation}
 \end{widetext}
where $x = \tanh \frac{u}{2}$ is the thermal saddle point solution and $X^\pm$, $Y^\pm$ parametrize the null shifts along the horizons due to the $SL(2,\mathbb{R})$ transformations. The $\theta$-functions indicate the contours on which the respective $SL(2,\mathbb{R})$ transformations have support (colored dashed lines in figure \ref{fold}). They are chosen such that the change in $SL(2,\mathbb{R})$ charges due to an operator pair has support along the part of the contour {\it between} the two operators \footnote{The precise form of the reparametrization $t(u)$ on contour 4 is not important, as it consists of just a single $SL(2,\mathbb{R})$ transformation, under which the Schwarzian action is invariant.}.

With this simplification, the path integral over repametrizations reduces to the integral over four variables $X^{\pm}$, $Y^{\pm}$. The Schwarzian action receives contributions from overlapping $SL(2,\mathbb{R})$ transformations on contour segments 3 and 5:
\begin{align}
    iS = -iC\int_{\text{contour}}du\ \{t(u),u\} = 2iC(X^+Y^-+Y^-Y^+)\,.
\end{align}

Performing the integral in \eqref{eq:F6SchwGen}, with appropriate normalizations we get the following integral representation of the correlator \eqref{correlator} for large $a$, $b$:
\begin{widetext}
\begin{align}
\label{integral}
\mathcal{F}_6 = \frac{1}{\Gamma(2\Delta_1)\Gamma(2\Delta_2)}\int_{0}^{+\infty} dp_1\int_0^{+\infty}dp_2\ p_1^{2\Delta_1-1}\ e^{-p_1}\ p_2^{2\Delta_2-1}\ e^{-p_2}\ \qty(\frac{1}{1+\frac{p_1p_2}{16C^2\delta_1\delta_2}e^{-\frac{2\pi}{\beta}(t_{wL}+t_{wR})}})^{2\Delta_j}\,,
\end{align}
\end{widetext}
where $p_{1,2}$ can be interpreted as the bulk momenta of the particles sourced by the operator insertions.
The dependence on $a$, $b$ has dropped out due to their being large.

We should note that the above approach is not as specific to $AdS_2$ gravity as it may seem. An {\it eikonal integral} similar to \eqref{integral} can be derived in higher dimensions by resumming exponentially enhanced gravitational contributions. The main difference will be additional dependence of the integrand on spatial directions (c.f., \cite{Shenker:2014cwa}).

\section{Result}

In the limit of large $\Delta_{1,2}$, we can perform the integral \eqref{integral} explicitly by saddle point approximation. Setting $p_1 = 2\Delta_1$ and $p_2 = 2\Delta_2$, we get
\begin{align}
\label{result_saddle}
    \mathcal{F}_6 \approx \qty(1+\frac{\Delta_1\Delta_2}{4C^2\delta_1\delta_2}e^{-\frac{2\pi}{\beta}(t_{wL}+t_{wR})})^{-2\Delta_{j}}
\end{align}
Note that this innocuous looking result is highly nontrivial from a microscopic point of view: it corresponds to the resummation of exponentially growing contributions from an infinite number of graviton exchange diagrams. It thus goes far beyond previous perturbative calculations of out-of-time-order six-point functions (e.g., \cite{Haehl:2017pak}).
It can be compared with the result of a bulk calculation in JT gravity: in the geodesic approximation, a two-point function between identical boundary operators is computed by $e^{-\Delta d}$ where $\Delta$ is the dimension of the operators and $d$ is the geodesic distance between the two boundaries. We obtain
\begin{align}
	\mathcal{F}_6 \approx \frac{e^{-\Delta \, d_\text{pert.} }}{e^{-\Delta\, d_\text{unpert.}}} = \ &\qty(1+\frac{1}{4}\frac{\delta S_1\delta S_2}{(S-S_0)^2} \, e^{-\frac{2\pi}{\beta}(t_{wL}+t_{wR})})^{-2\Delta_j}\label{collision_diagnosis}
\end{align}
Here, $\delta S_1$ and $\delta S_2$ are the changes of black hole entropy due to the two perturbations, and $S-S_0$ is the above-extremal entropy. 

We see that results \eqref{result_saddle} and \eqref{collision_diagnosis} match with each other once we identify $\frac{\delta S_i}{S-S_0}$ with $\frac{\Delta_i}{C\delta_i}$. 
From these expressions we can immediately infer the renormalized size of the operator creating the quantum circuit: 
\begin{widetext}
\begin{align}
\label{size_ren}
	\frac{n_{\text{ren}}}{n_{\text{max}}} = 1-\mathcal{F}_6 \approx \begin{cases}
	 0 &\quad\qquad\;\; -(t_{wL}+t_{wR})< 0\\
	 \frac{\Delta}{2}\frac{\delta S_1\delta S_2}{(S-S_0)^2} \,e^{-\frac{2\pi}{\beta}(t_{wL}+t_{wR})} &\quad\;\;\, 0\leq -(t_{wL}+t_{wR})<2t_*\\
		1-e^{-2\Delta\frac{2\pi}{\beta}\qty[-(t_{wL}+t_{wR})-2t_*]} &\quad 2t_*\leq -(t_{wL}+t_{wR})
	\end{cases}
\end{align}
\end{widetext}
where $2t_*$ is twice the scrambling time, $t_* \equiv \frac{\beta}{2\pi}\log( \frac{S-S_0}{\delta S})$.
The renormalized size \eqref{size_ren} only grows for negative $t_{wL}+t_{wR}$. This is expected as it corresponds to the presence of a collision. As we inject the perturbations earlier and $t_{wL}+t_{wR}$ becomes more negative, the size starts to grow exponentially in $-\frac{2\pi}{\beta}(t_{wL}+t_{wR})$. This is consistent with the fact the collision energy grows exponentially in $-\frac{2\pi}{\beta}(t_{wL}+t_{wR})$. When $-(t_{wL}+t_{wR})>2t_*$, the size saturates at its maximal value. On the gravity side, a large black hole has formed and the collision happens exponentially close to the singularity.\footnote{ The collision happens near the horizon of the unperturbed black hole, which has much smaller radius than the new black hole formed in the post-collision region. In other words, the singularity bends downward in the Penrose diagram.}

\section{Conclusion}

ER=EPR suggests a modern paradigm for understanding the time evolution of a two-sided black hole in terms of a quantum circuit shared between the boundaries. In this context, the collision in the interior of the wormhole can be interpreted as the overlap of two perturbations spreading through the quantum circuit. We diagnose this overlap by computing the size of the operator corresponding to the perturbed state. In this procedure, it appears that the existence of the singularity in the black hole interior plays a crucial role so that signals never meet if they are sent in too late. It will thus be interesting to understand what happens for geometries without a singularity. For instance, in charged black holes signals can always meet, but if they are sent late, the meeting will occur in a region where the dilaton is negative, i.e., gravity is repulsive. To understand this phenomenon remains an important open problem.

\begin{acknowledgments}
We thank Juan Maldacena, Xiaoliang Qi, Alex Streicher, Leonard Susskind for helpful discussions and comments. F.H.\ gratefully acknowledges support from the DOE grant DE-SC0009988 and from the Paul Dirac and Sivian Funds. Y.Z.\ is supported by the Simons foundation through the It from Qubit Collaboration.
\end{acknowledgments}

 \bibliographystyle{apsrev}

\end{document}